\documentclass[twocolumn,showpacs,preprintnumbers,amsmathamssymb]{revtex4}
\usepackage{graphicx}
\usepackage{color}
\usepackage{epsfig}
\preprint{graphite}
\begin{document}

\title{Universal conductivity and the electrodynamics of graphite at high pressures}

\author {A. Perucchi$^{1}$, L. Baldassarre$^{1}$, C. Marini$^{2,3}$, P. Postorino$^{2}$, F. Bernardini$^{4}$, S. Massidda$^{4}$ and S.~Lupi$^{2}$}
\affiliation{$^1$ Sincrotrone Trieste S.C.p.A., Area Science Park, I-34012 Basovizza, Trieste, Italy}\
\affiliation{$^2$ CNR-IOM  and and Dipartimento di Fisica, Universit\`a di Roma "La Sapienza", Piazzale Aldo Moro 2, I-00185 Roma, Italy}\
\affiliation{$^3$ European Synchrotron Radiation Facility, 6 Rue Jules Horowitz, BP220, 38043 Grenoble Cedex, France}\
\affiliation{$^4$ CNR-IOM Unit\`a SLACS and Dip. di Fisica, Universit\`a  di Cagliari, Cittadella Universitaria, I-09042 Monserrato (Ca), Italy}\
\date{\today}

\begin{abstract}

We address the in-plane pressure-dependent electrodynamics of graphite through synchrotron based infrared spectroscopy and {\it ab initio} Density Functional Theory calculations. The Drude term remarkably increases upon pressure application, as a consequence of an enhancement of both electron and hole charge densities. This is due to the growth of the band dispersion along the $k_z$ direction between the K and H points of the Brillouin zone. On the other hand, the mid-infrared optical conductivity between 800 and 5000 cm$^{-1}$ is almost flat, and very weakly pressure dependent, at least up to 7 GPa. This demonstrates a surprising robustness of the graphene-like universal quantum conductance of graphite, even when the interlayer distance is significantly reduced.

\end{abstract}


\date{\today}

\maketitle

There is no doubt in considering the discovery of graphene \cite{geim07} as one of the most important events in materials science, during the last decade. The reason for this lies not only on the incredible number of potential applications of this material, but also on the experimental observation of a completely new physics, which while already being theoretically predicted \cite{semenoff84}, was not believed to be realized in any existing material. Graphene amazes with such a richness of phenomena occurring in one of the most simple existing crystal structures.

Thanks to the discovery of graphene, also the research on its "real-world" analogue graphite, took momentum. Interestingly, many of the unusual phenomena discovered in graphene, such as the linear (Dirac-like) band dispersion at the H point \cite{zhou06}, the anomalous quantum Hall effect \cite{li07}, or the universal conductance properties \cite{kuzmenko08}, have been found to be present in graphite as well \cite{kopelevich07}. Understanding the transition from graphene to graphite is therefore of the uttermost importance. It is within this framework that one should address the role played by  the interlayer interaction in graphite, and ask what happens as long as the interlayer distance is externally modulated. The interlayer distance, for instance, can be expanded, by intercalating graphite with Br \cite{hwang11} or K \cite{gruneis08}. Alternatively one may try to study the behavior of graphite when the interlayer distance decreases and therefore the interaction among graphene layers enhances.

This is indeed the strategy we chose, by addressing the pressure-dependent electrodynamics of graphite with infrared reflectivity measurements and with new {\it ab initio} Density Functional Theory calculations. In particular, we aimed at noticing if pressure was able to demolish the universal  quantum conductivity of graphite, previously observed by Kuzmenko {\it et al.} \cite{kuzmenko08}.  As in the case of graphene, the optical conductivity of graphite at ambient pressure is ruled by transitions between hole and electron bands, yielding  a universal conductance $G_0=e^2/4\hbar$ per layer, which is almost frequency independent between 800 and 5000 cm$^{-1}$. 
As we will see in the following, pressure affects the out-of-plane band dispersion and the low-frequency conductivity of graphite by increasing the Drude spectral weight. Moreover, also at pressures as large as 7 GPa, a huge portion of the mid-infrared conductivity is practically independent of pressure. This shows that the universal quantum conductance of graphite is robust against a strong increase in the interlayer interaction.

A Highly Oriented Pyrolitic Graphite (HOPG) chunk (mosaic spread angle $<0.4^{\circ}$), purchased from SPI Supplies   of about 100 micron size was loaded into a screw-driven Diamond Anvil Cell (DAC). The sample was positioned in the DAC such as to obtain a flat and clean interface between the {\it ab} plane of graphite  and diamond. CsI salt was used as pressure transmitting medium, while a ruby chip loaded in the cell together with the sample is used as pressure calibrant through the fluorescence technique. Measurements have been performed with the help of a Bruker 66v spectrometer and Hyperion 2000 infrared microscope over an extended frequency range, between 350 and 13000 cm$^{-1}$. This was made possible by the use of synchrotron radiation \cite{lupi07}, as a bright source for far- and mid-infrared microspectroscopy.

\begin{figure}
\includegraphics[width=7.5cm]{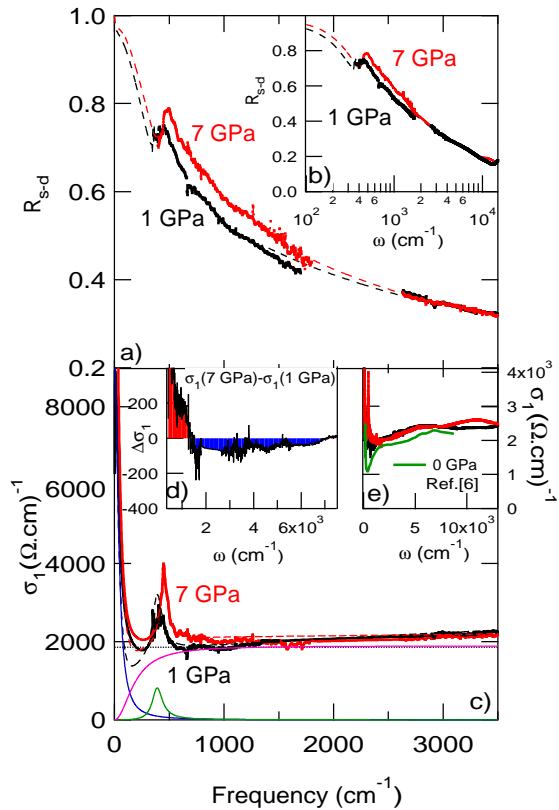}
\caption{Color online. a) Graphite reflectivity in the far and mid-infrared range at sample diamond interface at 1 and 7 GPa. Dashed lines show the extrapolated behavior, obtained from a Drude-Lorentz phenomenological fit, for Kramers-Kronig Transformations. b) Full range reflectivity at same pressures. c) Real part of the optical conductivity from Kramers-Kronig Transformations. The corresponding Drude-Lorentz fits (dashed lines), barely distinguishable from the Kramers-Kronig curves, are also plotted. Blue, green and pink curves correspond to the main components of the fit (see text). The black dotted line is a reference for the universal conductance G$_0$ expected for graphite (see Ref. \cite{kuzmenko08}). d) Difference of the optical conductivities between 1 and 7 GPa. e) Full range optical conductivity at 1 and 7 GPa, compared with ambient pressure data \cite{kuzmenko08}. }\label{graph1}
\end{figure}

The ambient temperature reflectivity of graphite basal plane at the sample-diamond interface ($R_{sd}(\omega)$) is reported in Fig. \ref{graph1}a. The break in the experimental data points between 1700 and 2500 cm$^{-1}$, is due to diamond multi-phonon absorptions. $R_{sd}(\omega)$ monotonically decreases with increasing frequency, except for a weak bump located at about 400 cm$^{-1}$. With increasing pressure, the slope of $R_{sd}(\omega)$ gets steeper in the spectral range below 3000 cm$^{-1}$, thus suggesting an increased metallicity, while the bump gets more pronounced and shifts towards high frequencies. Reflectivity data were first fitted within a Lorentz-Drude phenomenological framework \cite{wooten,dressel} (dashed lines in  Fig. \ref{graph1}a), and then extrapolated  in the missing ranges. This allowed to perform Kramers-Kronig Transformations, and calculate the real part of the optical conductivity $\sigma_1(\omega)$ (Fig. \ref{graph1}c), by following a procedure described elsewhere \cite{perucchi09}. The quality of the Drude-Lorentz fit can be verified in both panels a and c of Fig. \ref{graph1}. Fig. \ref{graph1}c also allows to visualize the main fitting components: one Drude term (blue), one narrow component at about 400 cm$^{-1}$ taking into account the bump in reflectivity (green), one strongly overdamped oscillator mimicking the graphene-like universal conductance (pink). A fourth component (not shown here) has also been employed to take into account electronic transitions in the near-infrared \cite{kuzmenko08,hanfland89}.

As previously found by Kuzmenko {\it et al.} \cite{kuzmenko08} at ambient conditions, the optical conductivity at 1 GPa of graphite is flat in a large portion of the infrared range, and its value per layer matches that of the frequency independent universal conductance $G_0=e^2/4\hbar$ of graphene. Deviations from G$_0$ are due to the presence of a narrow Drude term, and to a peak found at 400 cm$^{-1}$. On the low frequency side, the optical conductivity extrapolates at about 10000 ($\Omega$cm)$^{-1}$, 
in fair agreement with dc data \cite{edman98}. With increasing pressure the Drude term increases as well: This is the direct consequence of the $R_{sd}(\omega)$ enhancement described before. Moreover, the interband conductivity peak at 400 cm$^{-1}$ also increases and slightly blue-shifts. At larger frequencies, and up to the higher energy interband features  ($\geq$ 6000 cm$^{-1}$) \cite{hanfland89a}, the behavior of $\sigma_1(\omega)$ is the same as for graphene, even at pressures as high as 7 GPa. 
From the point of view of the optical spectral weight (SW), the pressure-dependent increase of the low-energy (Drude+peak) spectral weight is fully recovered at frequencies of the order of 8000 cm$^{-1}$, as shown in Fig. \ref{graph1}d, where the difference of the optical conductivities between 1 and 7 GPa is represented as a function of frequency.

\begin{figure}[thb]
\includegraphics[width=9cm]{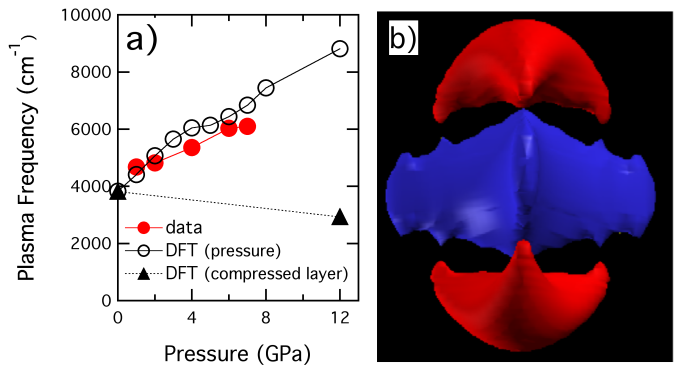}
\caption{Color online. a) Pressure dependent plasma frequency from data (solid symbols) and calculations (open symbols) plotted as a function of pressure. The plasma frequency calculated in the case of pure intralayer compression (no variation of interlayer parameters) is also shown (solid triangles).  b) Electron (blue) and holes (red) FS sheets in ambient pressure graphite.}\label{graph2}
\end{figure}

The most striking feature of our data is that while the overall graphite conductivity appears to be substantially unaffected by pressure, the Drude term enhancement is extremely significant. The in-plane Drude plasma frequency $\omega_p$, increases with pressure by a factor of 30\% (from 4700 to 6100 cm$^{-1}$), between 1 and 7 GPa (see red circles in Fig. \ref{graph2}a). Note that while we can assume a relatively large error bar of about 15\% in the absolute estimate of $\omega_p$ due to some arbitrariness in the Drude-Lorentz deconvolution, the relative error in the  $\omega_p$ pressure dependence is well below 5\%. On the other hand we know from literature \cite{hanfland89b} that between the same two pressure values, the intra-layer lattice parameter $a$ decreases by 0.5\% only (for comparison, the inter-layer lattice parameter $c$ decreases by 10\% in the same pressure range). This rules out the possibility that a bandmass reduction due to the intra-layer compression may be responsible for the observed increase of the plasma frequency. The enhancement of $\omega_p$ has rather to be ascribed to a variation in the number of the charge carriers upon pressure application. The microscopic mechanism underlying this effect will be investigated in the following, with the help of pressure-dependent {\it ab initio} calculations.

\begin{figure}[thb]
\includegraphics[width=9cm]{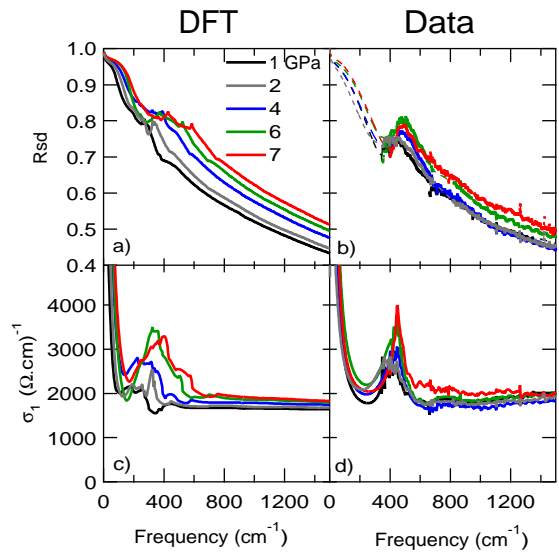}
\caption{Color online. Comparison of the pressure dependent optical data with the results of the {\it ab initio} calculation in terms of reflectivity at sample diamond interface (a,b) and optical conductivity (c,d).}\label{graph3}
\end{figure}

The present {\it ab initio} calculations are performed in the local density approximation to the density-functional theory (LDA-DFT)
as implemented in the all-electron LAPW code Wien2k \cite{Wien2k}.
A muffin tin radius of 1.3 Bohr is used, with a $R_{\rm MT} \times k_{max} =7$ plane wave cutoff.
Selfconsistency is achieved by  Brillouin Zone integration on a $60\times 60\times 20$
Mokhorst-Pack mesh. We use experimental lattice parameters according to 
Halfland {\it et al.}\cite{hanfland89b}.
The use of augmented plane waves has the advantages that wavefunction norm
conservation provides a direct access to the computation of the linear momentum operator $p_{ij}$ by the expectation
value of the gradient operator $\langle -i\hbar\nabla \rangle_{ij}$. This is especially useful for the determination of the
optical conductivity we compute as a sum of a intraband (Drude) and interband term. The Drude's term is computed from the
knowledge of the plasma frequencies $\omega_{p,\alpha}$ ($\alpha$ = $x(y)$ or $z$) defined as:
\begin{equation}
\label{omegap}
(\omega_{p,\alpha})^2= \frac{e^2}{\varepsilon_0 m_e^2}\int |p_{ii,\alpha}|^2\delta(\varepsilon_i(k)-E_{\rm F})\frac{d^3k}{(2\pi)^3}.
\end{equation}
For the interband term we calculate the (complex) dielectric function deriving the optical conductivity from the imaginary component given as:
\begin{equation}
\label{epsiloni}
\epsilon_{i,\alpha}(\omega) = \frac{1}{4\pi\varepsilon_0} \left(\frac{2\pi e}{m\omega}\right) \int |p_{ij,\alpha}|^2
                               \delta(\varepsilon_i(k)-\varepsilon_j(k)-\hbar\omega) \frac{d^3k}{(2\pi)^3},
\end{equation}

where $\varepsilon_j(k) < E_F < \varepsilon_i(k)$. 
The accuracy in the optical conductivity calculations comes from the correct description of the Fermi Surface (FS) topology and
convergence of the BZ summation in Eqs. (\ref{omegap}) and (\ref{epsiloni}). As for the FS topology
we obtain for the ambient pressure graphite two sheets (electrons and holes) shown in Fig. \ref{graph3}b. The shape for the electron (blue)
and holes (red) is in agreement with the results of Ref.\onlinecite{dresselhaus74} and \onlinecite{Rubio2008}.
The convergence of the BZ summation is very important in computing the plasma frequency. To get a confidence within a few
percent we found necessary to go up to a mesh equivalent to a $600\times600\times100$ Monkhorst-Pack.

Knowledge of the complex dielectric function allows to recalculate the reflectivity at the sample diamond interface, at any given pressure.  As reported in Fig. \ref{graph3}, it can be easily seen that the overall agreement between the calculations of $R_{sd}(\omega)$ and $\sigma_1(\omega)$ with data is good. 
The plasma frequency values, both from data and calculation, are reported in Fig. \ref{graph2}a, as a function of pressure. 
It is immediately seen that the steep enhancement of $\omega_{p}$ described previously, is confirmed by calculations. 
As outlined above, this effect can not be attributed to a bandwidth increase (i.e. a mass reduction) due to the slight in-plane compression. 
To validate this hypothesis we have performed {\it ab initio} calculation for a fictitious graphite system, subject to an in-plane compression, but with the same interlayer spacing as ambient pressure graphite. The calculation (triangles in Fig. \ref{graph2}a) shows that under these conditions $\omega_{p}$ does not increase upon intra-layer compression, but even slightly decreases. This tells us that the intra-layer compression induced by pressure has only a negligible effect on the in-plane electronic properties. 

\begin{figure}[thb]
\includegraphics[width=8cm]{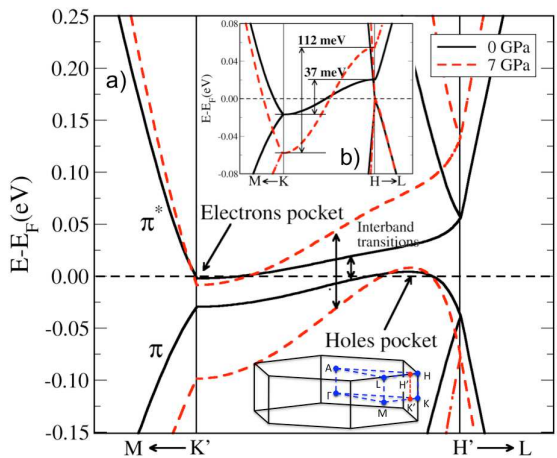}
\caption{Color online. Graphite band structure dispersion at 0 and 7 GPa from {\it ab initio} calculations. a) Band structure along the M-K'-H'-L directions (see text). The figure highlights the mechanisms responsible for the pressure dependent increase of carrier concentration, as well as for the blue shift of the interband absorption peak experimentally observed at about 400 cm$^{-1}$ (see text). b) Same along the high symmetry lines M-K-H-L, where the bonding $\pi$ and antibonding $\pi^*$ bands get degenerate.}\label{graph4}
\end{figure}

In Fig. \ref{graph4}a we show the band structure dispersion for free standing and
compressed (7 GPa) graphite along the
M-K'-H'-L directions, where K'-H' stands for a line parallel to the K-H but
slightly shifted towards the M point (see lower inset). This choice allows to see more clearly the bonding $\pi$ and antibonding $\pi^*$ bands that are degenerate along the KH direction (see upper inset). 
The Figure shows the origin of the holes and electrons Fermi surfaces sheets of Fig. \ref{graph2}b
from the $\pi$ and $\pi^*$ bands respectively.  
Pressure broadens the bands along the $k_z$ direction between points K' and H' in Fig. \ref{graph4}a, 
while leaving the in-plane dispersion (along K'M and H'L) almost unchanged. The increased $k_z$ dispersion moves the K and H semi-metallic points further apart in energy, passing from 37 meV at ambient pressure to 112 meV at 7 GPa, with the chemical potential lying in between. 
This results at K and H, in a relative shift - of opposite sign - of the semi-metallic crossings with respect to the chemical potential, thus leading to an increased number of free charge carriers of both signs (Fig. \ref{graph4}b). This qualitative assessment is confirmed by a more detailed calculation showing how the number of free electrons and holes increases with pressure by the same amount. 
Moreover compression is also responsible for the shift of the peak around 400 cm$^{-1}$ in the optical conductivity. Indeed this peak results from the interband transitions shown in Fig. \ref{graph4}a (vertical arrows). Along the KH direction this absorption does vanish because of the degeneracy of $\pi$ and $\pi^*$ bands. In the neighboroud of that direction (namely K'H') the bands are non degenerate and their contribution reaches its maximum value, thus giving rise to a peak in $\sigma_1(\omega)$, where $\pi$ and $\pi^*$ are full and empty respectively.  The direct gap along the K'H' direction strongly depends on interlayer distance and justifies the upshift of the peak in the optical conductivity as experimentally observed.

As a final remark, we comment  on the pressure-induced spectral weight redistribution from the far to the mid-infrared, as shown in Fig. \ref{graph1}d. Interestingly, this behavior is qualitatively similar to that observed by Horng {\it et al.} \cite{horng10} in biased graphene. These authors find a transfer of spectral weight from low ($\leq$ 1000 cm$^{-1}$) to high ($\geq$ 1500 cm$^{-1}$) frequencies. In graphene this behavior is explained by the electric-field tuning of the chemical potential close to the Dirac point: When the chemical potential moves away from the Dirac point, the Fermi Surface (and thus the Drude term) increases at the expense of the low energy interband transitions \cite{kuzmenko08}. 
In a similar way, pressure in graphite, increases the number of charge carriers, while depleting the low energy interband transitions following the same mechanisms as described in Refs. \onlinecite{kuzmenko08,horng10} and \onlinecite{li08}. To this aim, the only band-structure element needed is the presence of a band crossing close to the chemical potential, independent from the details of the band dispersion (either conic as in graphene or parabolic as in graphite K point).

We have shown in this letter how pressure affects the in-plane low energy electrodynamics of graphite, by increasing the number of both electron and hole charge carriers. This effect is due to a pressure-induced shift of opposite sign, of the bands at the K and H point, with respect to the chemical potential. The other effect of pressure is the enhancement and blue shift of an inter-band transition located at about 400 cm$^{-1}$. Above this frequency, the effects of pressure on the in-plane optical conductivity are rather limited, at least up to 1 eV, where the inter-band transitions previously studied by Hanfland {\it et al.} \cite{hanfland89b} take place. Nonetheless, a very large portion of the mid-infrared spectral range between 800 and 5000 cm$^{-1}$ is very weakly pressure-dependent, thus demonstrating a surprising robustness of the universal conductance properties of graphite when the interlayer distance is significantly reduced.

F.B. acknowledges support from CASPUR under the Standard HPC Grant 2012.


\begin{thebibliography}{99}
\bibitem{hanfland89a} The pressure dependence of the near-infrared interband features above 6000 cm$^{-1}$ has already been addressed by \cite{hanfland89}, and will not be discussed in detail here.
\end{thebibliography}
\end{document}